\documentclass{ifacconf}
\usepackage{natbib}
\usepackage{url}

\usepackage{amsmath,amssymb,mathtools}

\usepackage{graphicx}
\usepackage{float}

\usepackage{algorithm}
\usepackage{algcompatible}
\usepackage{multirow} 
\usepackage[table]{xcolor}
\usepackage{array}
\usepackage{tabularx}
\usepackage{caption}

\newcolumntype{L}[1]{>{\raggedright\arraybackslash}p{#1}}
\newcolumntype{C}[1]{>{\centering\arraybackslash}p{#1}}

\newcommand{\R}{\mathbb{R}}

\newcommand{\re}{\operatorname{Re}}

\newcommand{\imid}{\operatorname{midpoint}}

\newcommand{\Kraw}{\mathcal{K}}

\begin{document}
\begin{frontmatter}
\title{
Certified Detection of Bifurcation Candidates in Uncertain Nonlinear Systems using Interval Analysis
}
\author[First]{R. Prakash} 
\author[First]{S. Janardhanan} 
\author[First]{S. Sen}
\address[First]{Department of Electrical Engineering, Indian Institute of Technology Delhi, New Delhi--110016, India (e-mail: rudra.prakash@ee.iitd.ac.in, janas@ee.iitd.ac.in, shaunak.sen@ee.iitd.ac.in).}

\begin{abstract}
Qualitative transitions in nonlinear dynamical systems (e.g., loss of stability, onset of oscillations, emergence of multistability) delimit operating regimes and can arise as implicit constraints in robust analysis and design under parametric uncertainty. When parameters are inferred from data, admissible values are naturally represented as uncertainty sets, motivating certified tests for the presence or absence of regime-transition candidates. We propose a validated interval workflow that encodes saddle-node and Hopf \emph{candidate} conditions as square augmented algebraic systems and applies the Krawczyk operator to certify, over a prescribed state--parameter box, either (i) existence and local uniqueness of a candidate solution or (ii) certified absence. Numerical experiments on uncertain synthetic gene-network ODE models yield locally certified saddle-node candidate enclosures on a two-parameter slice for a bistable circuit and certified Hopf candidate enclosures for a three-state oscillator using a Routh--Hurwitz specialization. The resulting certificates are intended to support regime-aware analysis and design under bounded uncertainty by complementing non-validated, pointwise baselines (e.g., Newton method solves at discrete parameter values) and sampling-based workflows with rigorous presence/absence guarantees on user-specified parameter slices.
\end{abstract}


\end{frontmatter}

\section{Introduction}
Many control and decision-making tasks depend not only on quantitative performance indices but also on qualitative properties of nonlinear dynamical systems \citep{khalil_nonlinear}. In dynamic optimization and optimal control, stability margins, oscillation onset, and multistability can appear as implicit feasibility constraints that delimit admissible operating regimes \citep{bertsekas_dpoc,rawlings_mpc}. In data-driven settings, these constraints must be enforced despite model uncertainty arising from state/parameter estimation and system identification.

Bifurcation analysis formalizes such qualitative transitions by relating stability changes to parameter variation and by tracking equilibria or limit cycles and their local spectra \citep{strogatz_nonlinear,kuznetsov_bifurcation,seydel_bifurcation}. Classical workflows rely on numerical continuation and eigenvalue-based stability assessment at fixed parameter values \citep{kuznetsov_bifurcation,seydel_bifurcation}. While effective for nominal models, these computations are non-validated and provide no direct guarantee as parameters range over uncertainty sets. Under uncertainty, Monte Carlo sampling combined with repeated continuation or simulation is commonly used; however, sampling provides no coverage guarantees, can miss thin transition regions, and does not yield certificates of absence. These limitations are especially problematic when regime boundaries are embedded in higher-level iterative workflows, such as robust analysis and design under parametric uncertainty and optimal experiment design for model discrimination near transitions. To contextualize the proposed certificates, we report non-validated numerical reference loci---computed using pointwise multistart Newton solves at discrete parameter values on the same parameter slices---alongside the validated enclosures in the case studies.

Validated numerical methods, in particular interval analysis, provide rigorous alternatives for nonlinear equations and dynamical systems \citep{moore_interval,jaulin_applied_interval,tucker_validated,alefeld1994verification,moore1977test,moore1980generalized,chorasiya2023quantitative,prakash2025rigorous}. Interval-Newton and Krawczyk operators can certify existence and uniqueness (or exclusion) of solutions of nonlinear systems over a box \citep{hsu_krawczyk,rump1982solving}, providing sound feasibility tests under bounded uncertainty.

In this paper, we develop a validated interval workflow for certifying saddle-node and Hopf \emph{candidate} conditions in uncertain ODE models by encoding these conditions as square augmented algebraic systems and applying Krawczyk inclusion/exclusion tests. Although the methodology is broadly applicable to nonlinear systems with interval-bounded parameters, we focus on a motivating application from synthetic biology: gene-network models whose kinetic parameters are typically estimated with substantial uncertainty and whose desired behaviors are organized by saddle-node and Hopf transitions. Synthetic gene networks exhibit multistability (toggle switches) and oscillations (repressilator-type circuits), making them natural benchmarks for control-relevant regime certification \citep{gardner_toggle,elowitz_repressilator,stricker_oscillator}.

The main contributions are:
\begin{itemize}
  \item A square augmented-equation formulation of saddle-node and Hopf \emph{candidate} conditions that is directly amenable to interval-based inclusion/exclusion tests under bounded parametric uncertainty.
  \item A Krawczyk-operator-based certification result that, for a given state--parameter box, returns either (i) a validated enclosure containing a unique solution of the candidate equations or (ii) a validated exclusion proving that no candidate solution exists in the box.
  \item A hybrid search algorithm—incorporating numerical seeding, local boxing strategies, Krawczyk-based inclusion–exclusion tests, and interval bisection—is employed to construct certified box collections, which are subsequently aggregated into candidate sets defined on user-specified parameter slices.
  \item Two case studies on uncertain nonlinear ODE models of synthetic gene networks that demonstrate certified saddle-node and Hopf candidate enclosures, including a three-state specialization based on Routh--Hurwitz conditions.
\end{itemize}

Section~\ref{sec:prelim} introduces interval arithmetic and the Krawczyk operator; Section~\ref{sec:problem} formulates saddle-node and Hopf \emph{candidate} systems; Section~\ref{sec:guarantees} states the validation guarantees and presents the Interval Candidate Finder search procedure (Section~\ref{sec:algorithm}); Section~\ref{sec:experiments} reports numerical experiments and certified enclosures; Sections~\ref{sec:discussion}--\ref{sec:conclusion} discuss limitations and conclude.

\section{Preliminaries}\label{sec:prelim}
\subsection{Interval arithmetic}\label{sec:interval_arith}
An interval $[a]=[\underline a,\overline a]\subset\R$ encloses all real numbers between its bounds. Its midpoint is $\imid([a])\coloneqq (\underline a+\overline a)/2$. Interval vectors (boxes) $[x]=[x_1]\times\cdots\times[x_n]\subset\R^n$ and interval matrices $[A]\subset\R^{n\times n}$ are defined elementwise \citep{moore_interval,neumaier_interval,jaulin_applied_interval}; for a box $[z]$, the midpoint $\imid([z])$ is taken componentwise. For a function $g$, an \emph{interval extension} $[g]([z])$ is any enclosure satisfying $g(z)\in[g]([z])$ for all $z\in[z]$; in validated computations, outward rounding is used to preserve this inclusion property.

\subsection{Krawczyk operator}\label{sec:krawczyk}
Consider a continuously differentiable mapping $g:\R^m\to\R^m$ and a search box $[z]$. Let $z_c=\imid([z])$ be its midpoint and let $C\in\R^{m\times m}$ be a nonsingular preconditioner, typically an approximation of $(\nabla g(z_c))^{-1}$. The Krawczyk operator \citep{hsu_krawczyk,jaulin_applied_interval, neumaier_interval} is defined by
\begin{equation}
  \Kraw([z]) \;=\; z_c - C\,g(z_c) + \bigl(I - C\,[\nabla g]([z])\bigr)([z]-z_c).
\end{equation}
It provides two certification tests \citep{neumaier_interval}:
\begin{itemize}
  \item \emph{Existence and uniqueness:} if $\Kraw([z])\subset \operatorname{int}([z])$, then there exists a unique $z^*\in[z]$ such that $g(z^*)=0$.
  \item \emph{Exclusion:} if $\Kraw([z])\cap[z]=\emptyset$, then $g(z)=0$ has no solution in $[z]$.
\end{itemize}
These tests form the basis of hybrid validated search procedures that combine interval bisection with Krawczyk-based inclusion--exclusion pruning to return enclosures of unique solutions and certificates of absence for nonlinear equation systems.

\section{Problem formulation}\label{sec:problem}
We consider a parametric nonlinear ODE model
\begin{equation}
  \dot x = f(x,\theta), \qquad x\in\R^n,\; \theta\in\Theta\subset\R^p,
\end{equation}
where $f$ is continuously differentiable and the admissible parameter set $\Theta$ is an interval box capturing parametric uncertainty. The objective is to certify, for prescribed state--parameter boxes, whether augmented algebraic systems encoding saddle-node or Hopf \emph{candidate} conditions (i) admit a locally unique solution or (ii) admit no solution.

A steady state $x^*$ satisfies $f(x^*,\theta)=0$. Let $J(x,\theta)=\nabla_x f(x,\theta)$ denote the Jacobian with respect to the state.

\subsection{Saddle-node candidate condition}
A saddle-node \emph{candidate} at a steady state satisfies
\begin{equation}
  f(x,\theta)=0,\qquad \det\bigl(J(x,\theta)\bigr)=0,
  \label{eq:sn_det}
\end{equation}
\citep{kuznetsov_bifurcation,seydel_bifurcation}. This determinant condition is convenient for defining a square augmented system; in applications, it may be evaluated with interval matrix operations or replaced by analytically reduced equivalent conditions when available.

For certified search on a parameter slice, we designate one scalar component of $\theta$ as the continuation (bifurcation) parameter $\mu$ and collect the remaining parameters into $\vartheta$. The saddle-node candidate equations are written as the square augmented system
\begin{equation}
  g_{\mathrm{SN}}(x,\mu;\vartheta)=\begin{bmatrix} f(x,\mu,\vartheta)\\ \det\bigl(J(x,\mu,\vartheta)\bigr)\end{bmatrix}=0,
  \label{eq:sn_square}
\end{equation}
which is directly amenable to validated inclusion/exclusion via interval-Newton or Krawczyk operators.

\subsection{Hopf candidate condition}
A Hopf \emph{candidate} at a steady state satisfies an eigenvalue condition in which $J(x,\theta)$ has a conjugate pair of purely imaginary eigenvalues $\lambda_{1,2}=\pm j\omega$ with $\omega>0$ \citep{kuznetsov_bifurcation,guckenheimer_holmes,seydel_bifurcation}. A determinant-based characterization is obtained by introducing $\omega$ and enforcing
\begin{equation}
  f(x,\theta)=0,\qquad \det\bigl(J(x,\theta)-j\omega I\bigr)=0.
  \label{eq:hopf_det}
\end{equation}
This form is convenient for defining a square augmented system; in applications, it may be evaluated using interval extensions and/or replaced by analytically reduced real-valued conditions on restricted branches. For comparison, non-validated numerical routines for directly computing Hopf points are classical in the continuation literature \citep{roose1985numerical}.

\textit{Specialization to three-state models.} For three-state circuits, Hopf candidates can be detected without eigenvector variables using Routh--Hurwitz conditions for the characteristic polynomial $\lambda^3+a_1\lambda^2+a_2\lambda+a_3=0$ \citep{kuznetsov_bifurcation,seydel_bifurcation}. Defining
\begin{equation}
  a_1=-\operatorname{trace}(J),\qquad a_2=\sum_{i<j} \det(J_{ij}),\qquad a_3=-\det(J),
  \label{eq:rh_coeffs}
\end{equation}
where $J_{ij}$ denotes the $2\times2$ principal submatrix indexed by $\{i,j\}$, a Hopf \emph{candidate} in this setting satisfies
\begin{equation}
  f(x,\theta)=0,\qquad a_1>0,\; a_2>0,\; a_3>0,\qquad a_1a_2-a_3=0.
  \label{eq:hopf_rh}
\end{equation}
The trace and determinant quantities above can be computed using interval matrix operations, yielding a purely algebraic test compatible with validated inclusion/exclusion. In particular, in the three-state specialization we enforce the strict Routh--Hurwitz inequalities by interval sign checks (e.g., verifying that the enclosures of $a_1,a_2,a_3$ are strictly positive on the accepted boxes) in addition to certifying the equality constraint $a_1a_2-a_3=0$ via the Krawczyk inclusion test.

\section{Theoretical results}\label{sec:guarantees}
We summarize the validated inclusion and exclusion tests used to certify solutions of the augmented algebraic systems encoding saddle-node and Hopf \emph{candidate} conditions.

\subsection{Validated candidate certification via Krawczyk}\label{sec:cert_boxes}
Let $g$ denote a square augmented algebraic system encoding a saddle-node or Hopf \emph{candidate} condition (e.g., $g\equiv g_{\mathrm{SN}}$ in \eqref{eq:sn_square} or $g\equiv g_{\mathrm{H}}$ in \eqref{eq:hopf_det} or \eqref{eq:hopf_rh}). For a candidate box $[z]$ in the augmented variables, we evaluate the Krawczyk operator (Section~\ref{sec:krawczyk}) using an interval enclosure of $[\nabla g]([z])$.

\textbf{Theorem~1 (Validated candidate enclosure).}
Let $[z]\subset\R^m$ be a box for the augmented unknowns $z$. If $\Kraw([z])\subset\operatorname{int}([z])$, then the augmented system $g(z)=0$ has a unique solution $z^*\in[z]$. Writing $z=(x,\theta,\eta)$ (state, parameters, and auxiliary variables), the projections $x^*\in[x]$ and $\theta^*\in[\theta]$ yield an equilibrium of the original ODE and satisfy the selected saddle-node or Hopf \emph{candidate} algebraic conditions at $(x^*,\theta^*)$.

\emph{Proof.} The inclusion $\Kraw([z])\subset\operatorname{int}([z])$ implies existence and uniqueness of a zero of the square nonlinear system $g(z)=0$ in $[z]$ by the standard Krawczyk inclusion principle. The interpretation of the components of $z^*$ for \eqref{eq:sn_square} and the Hopf formulations \eqref{eq:hopf_det}--\eqref{eq:hopf_rh} is immediate. Under standard nondegeneracy (genericity/transversality) assumptions \citep{kuznetsov_bifurcation,seydel_bifurcation}, these certified algebraic conditions correspond to a local saddle-node or Hopf bifurcation; in this work we do not validate the remaining bifurcation-theorem hypotheses such as transversality of the critical eigenvalue crossing, isolation of the critical eigenvalues from the imaginary axis (for Hopf), or higher-order nondegeneracy conditions (e.g., fold coefficient / first Lyapunov coefficient), and therefore we report \emph{candidate} enclosures. \hfill$\square$

\emph{Remark (standard nondegeneracy conditions).} For a saddle-node (fold) at $(x^*,\mu^*)$, one typically assumes that $J(x^*,\mu^*)$ has a simple zero eigenvalue (equivalently, $\dim\ker(J(x^*,\mu^*$ $))=1$) with right/left nullvectors $v,w$ normalized so that $w^\top v=1$, together with fold nondegeneracy $w^\top f_{xx}(x^*,\mu^*)[v,v]\neq 0$ and parameter transversality $w^\top f_{\mu}(x^*,\mu^*)\neq 0$. For a Hopf bifurcation at $(x^*,\mu^*)$, one typically assumes a simple conjugate pair $\pm j\omega^*$ with $\omega^*>0$ (algebraic multiplicity one), all other eigenvalues strictly off the imaginary axis, and a transverse crossing, i.e., $\frac{d}{d\mu}\re\lambda(\mu)\big|_{\mu=\mu^*}\neq 0$ for the critical pair; a generic Hopf further assumes a nonzero first Lyapunov coefficient.

\textbf{Theorem~2 (Validated exclusion for a box).}
If $\Kraw([z])\cap[z]=\emptyset$, then the augmented equations have no solution in $[z]$. Consequently, there is no saddle-node or Hopf \emph{candidate} consistent with the projected state/parameter sub-boxes $[x]$ and $[\theta]$ associated with $[z]$.

\emph{Proof.} This is the standard Krawczyk exclusion property. \hfill$\square$

\emph{Remark (soundness).} For a given box $[z]$, Theorem~1 provides a validated enclosure of a unique solution of $g(z)=0$ within $[z]$, while Theorem~2 provides a validated certificate that no solution exists in $[z]$. In a domain-covering hybrid search (interval bisection coupled with validated, Krawczyk-based inclusion--exclusion pruning), any remaining regions not certified by either test must be tracked explicitly as undecided.

\subsection{Interval Candidate Finder}\label{sec:algorithm}
We employ a hybrid search procedure over an initial augmented box $[z_0]$ to isolate boxes that satisfy the validated inclusion test (Theorem~1) and to discard boxes that satisfy the validated exclusion test (Theorem~2).

\begin{algorithm}[h!]
\caption{Interval Candidate Finder (saddle-node or Hopf)}
\label{alg:ibf}
\begin{algorithmic}[1]
\STATE \textbf{Input:} ODE $f(x,\theta)$; admissible parameter box $\Theta$; initial state box $[x_0]$; tolerance $\varepsilon$.
\STATE \textbf{Select test:} saddle-node system $g\equiv g_{\mathrm{SN}}$ in \eqref{eq:sn_square} or Hopf system $g\equiv g_{\mathrm{H}}$ in \eqref{eq:hopf_det} (or \eqref{eq:hopf_rh} for three-state models).
\STATE \textbf{Augmented variables:} define $z$ as the vector of unknowns for the selected system (state, auxiliary variables, and any parameters treated as decision variables on the chosen slice, e.g., $\mu$ and possibly $\omega$ for Hopf), fix the remaining parameters within their prescribed intervals, and set $[z_0]$ accordingly.
\STATE \textbf{Width criterion:} let $\operatorname{width}([z])=\max_i\,(\overline z_i-\underline z_i)$.
\STATE \textbf{Initialize:} worklist $\mathcal{W}\leftarrow\{[z_0]\}$; validated list $\mathcal{S}\leftarrow\emptyset$; undecided list $\mathcal{U}\leftarrow\emptyset$.
\WHILE{$\mathcal{W}\neq\emptyset$}
  \STATE Remove a box $[z]$ from $\mathcal{W}$.
  \STATE Compute $\Kraw([z])$ for $g(z)=0$.
  \IF{strict inequalities are required (e.g., \eqref{eq:hopf_rh}); compute the required $[a_i]$ on $[z]$, and if any has $\overline a_i\le 0$}
    \STATE Discard $[z]$.
  \ELSIF{$\Kraw([z])\subset\operatorname{int}([z])$ \AND (no strict inequalities are required, or all required $\underline a_i>0$)}
    \STATE Add $[z]$ to $\mathcal{S}$.
  \ELSIF{$\Kraw([z])\cap[z]=\emptyset$}
    \STATE Discard $[z]$ (validated exclusion of solutions in $[z]$).
  \ELSIF{$\operatorname{width}([z])\le\varepsilon$}
    \STATE Add $[z]$ to $\mathcal{U}$.
  \ELSE
    \STATE Choose an index $i\in\arg\max_j (\overline z_j-\underline z_j)$.
    \STATE Bisect $[z]$ at the midpoint along coordinate $i$ and add both children to $\mathcal{W}$.
  \ENDIF
\ENDWHILE
\STATE \textbf{Output:} validated boxes $\mathcal{S}$ and undecided boxes $\mathcal{U}$.
\end{algorithmic}
\end{algorithm}

\emph{Practical considerations.} Tight interval extensions and additional contractors (e.g., constraint propagation for $f(x,\theta)=0$) can reduce bisection. Certification is box-local and improves as $\varepsilon$ decreases.

\section{Numerical experiments}\label{sec:experiments}
This section reports numerical experiments that certify saddle-node and Hopf \emph{candidate} conditions on selected parameter slices for uncertain nonlinear ODE models. All computations were performed in \texttt{Julia}~1.11.0 using \texttt{IntervalArithmetic.jl}~0.20.9.

\subsubsection{Non-validated numerical baselines.}
To facilitate comparison with standard workflows, we additionally compute non-validated numerical reference loci on the same parameter slices as the validated enclosures. Specifically, we compute discrete reference saddle-node and Hopf loci via pointwise multistart Newton solves on a grid of parameter values for the corresponding augmented candidate systems, seeding Newton from multiple initial guesses. We also perform coarse parameter sampling on the selected slices, using equilibrium solves and local linear stability checks (for fixed points) and/or long-horizon time simulations (for oscillatory behavior), to visualize qualitative regime changes across the certified candidate regions. These baseline computations are intended only as qualitative reference and are not used to justify the validation guarantees.
\subsection{Toggle-switch and saddle-node candidate certification}
We consider a two-gene toggle switch with Hill repression \citep{gardner_toggle}:
\begin{equation}
\begin{aligned}
  \dot x_1 &= \frac{\alpha_1}{1 + x_2^{n_2}} - \delta_1 x_1; \quad
  \dot x_2 &= \frac{\alpha_2}{1 + x_1^{n_1}} - \delta_2 x_2.
\end{aligned}
\label{eq:toggle}
\end{equation}
with state $x=(x_1,x_2)\in\R_{\ge 0}^2$ and uncertain parameters
\(
  \Theta:\;
  \alpha_1\in[5,15],\,
  n_1\in[1.5,2.5],\,
  (\alpha_2,\delta_1,\delta_2,n_2)=(12,1.2,0.8,1.6).
\)
The model exhibits bistability via saddle-node bifurcations for sufficiently strong nonlinearity.

We search for saddle-node candidates by applying Algorithm~\ref{alg:ibf} to \eqref{eq:sn_square}, selecting $\mu=\alpha_1$ as the bifurcation parameter and fixing $(\alpha_2,\delta_1,\delta_2,n_2)=(12,1.2,0.8,1.6)$ while varying $n_1\in[1.5,2.5]$. The output is a set of locally certified candidate boxes in $(x,\mu)$ whose projection onto $(\alpha_1,n_1)$ summarizes candidate locations on the selected slice.
We report the certified saddle-node candidate enclosures obtained for the toggle-switch model.
\subsubsection{Certified saddle-node candidate strips and projection.}
Table~\ref{tab:recommended_slice_sn_strips} lists locally validated saddle-node \emph{candidate} strips indexed by $n_1$. Each row reports validated enclosures of the equilibrium state $(x_1,x_2)$ and the corresponding interval for the reconstructed parameter value $\alpha_1$. Due to space constraints, we report only a representative subset of the computed enclosures.

\begin{figure}[t!]
\centering
\includegraphics[width=0.75\linewidth]{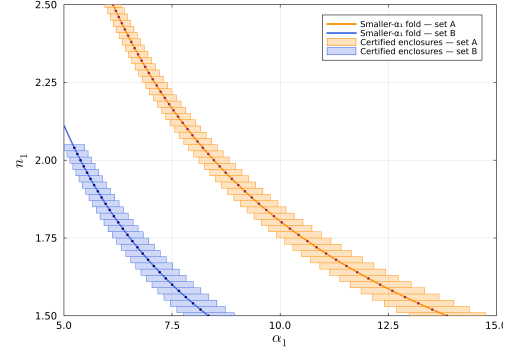}
\caption{Effect of the fixed parameters on the smaller-$\alpha_1$ saddle-node fold of the toggle-switch model on the $(n_1,\alpha_1)$ slice. Certified enclosures use $(\alpha_2,\delta_1,\delta_2,n_2)=(12,1.2,0.8,1.6)$ and vary $(\alpha_1,n_1)$. For comparison, set B uses $(\alpha_2,\delta_1,\delta_2,n_2)=(10,1,1,2)$ to illustrate sensitivity. The solid orange and dashed blue curves show the numerically computed smaller-$\alpha_1$ fold loci for sets A and B, respectively. Shaded rectangles indicate interval-certified saddle-node candidate enclosures (Table~\ref{tab:recommended_slice_sn_strips}), and markers denote enclosure midpoints.}
\label{fig:bifbox}
\end{figure}

\begin{figure}[t!]
\centering
\includegraphics[width=0.6\linewidth]{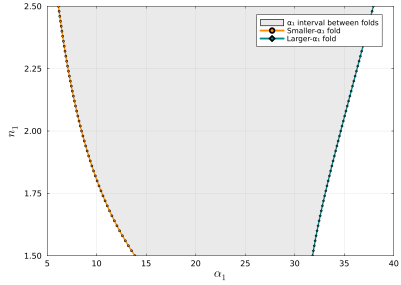}
\caption{Toggle-switch model on the $(n_1,\alpha_1)$ slice (set A, $(\alpha_2,\delta_1,\delta_2,n_2)=(12,1.2,0.8,1.6)$): numerical smaller-$\alpha_1$ (solid orange) and larger-$\alpha_1$ (dashed teal) saddle-node fold loci; shading shows the intervening $\alpha_1$ interval. Interval-certified enclosures are reported separately.}
\label{fig:smaller_larger_alpha1_folds}
\end{figure}

\emph{Remark.} In a bistable regime, a fixed-parameter slice is typically delimited by two saddle-node (fold) candidates at a given $n_1$, corresponding to a lower fold (smaller $\alpha_1$) and an upper fold (larger $\alpha_1$). Table~\ref{tab:recommended_slice_sn_strips} reports at most one locally validated candidate per sampled $n_1$, determined by the chosen seeding strategy and local search window. Identification and certification of both folds can be obtained by using multiple seeds and/or a domain-covering search over the relevant state--parameter region.
\begin{table*}[t!]
\centering
\caption{Toggle-switch case study: locally validated saddle-node \emph{candidate} strips obtained by applying the Krawczyk inclusion test to the saddle-node candidate equations on a selected parameter slice. Each row reports validated interval enclosures of the equilibrium state $(x_1,x_2)$ and the corresponding reconstructed interval for the parameter $\alpha_1$ at the sampled Hill coefficient $n_1$ (the displayed $n_1$ interval indicates the sampling strip).}
\label{tab:recommended_slice_sn_strips}
\setlength{\tabcolsep}{10pt}
\begin{tabular}{|c|c|c|c|c|c|c|}
\hline
\rowcolor{blue!10}
\textbf{$n_1$} & \textbf{$n_1$ interval} & \textbf{$x_1$ interval} & \textbf{$x_2$ interval} & \textbf{$\alpha_1$ interval} & \textbf{midpoint state} & \textbf{midpoint $\alpha_1$} \\
\hline
1.76 & $[1.75, 1.77]$ & $[5.2822, 5.6097]$ & $[0.6993, 0.7467]$ & $[9.9152, 10.9501]$ & $(5.4460, 0.7230)$ & 10.4245 \\
\hline
1.78 & $[1.77, 1.79]$ & $[5.2213, 5.5399]$ & $[0.6912, 0.7378]$ & $[9.7358, 10.7345]$ & $(5.3806, 0.7145)$ & 10.2274 \\
\hline
1.8 & $[1.79, 1.81]$ & $[5.1617, 5.4719]$ & $[0.6834, 0.7291]$ & $[9.5628, 10.5271]$ & $(5.3168, 0.7063)$ & 10.0376 \\
\hline
1.82 & $[1.81, 1.83]$ & $[5.1033, 5.4054]$ & $[0.6758, 0.7207]$ & $[9.3958, 10.3275]$ & $(5.2544, 0.6983)$ & 9.8546 \\
\hline
1.84 & $[1.83, 1.85]$ & $[5.0463, 5.3406]$ & $[0.6685, 0.7126]$ & $[9.2345, 10.1352]$ & $(5.1934, 0.6905)$ & 9.6782 \\
\hline
1.86 & $[1.85, 1.87]$ & $[4.9905, 5.2773]$ & $[0.6613, 0.7047]$ & $[9.0787, 9.9499]$ & $(5.1339, 0.6830)$ & 9.5079 \\
\hline
1.88 & $[1.87, 1.89]$ & $[4.9359, 5.2155]$ & $[0.6543, 0.6970]$ & $[8.9281, 9.7713]$ & $(5.0757, 0.6757)$ & 9.3436 \\
\hline
1.9 & $[1.89, 1.91]$ & $[4.8825, 5.1551]$ & $[0.6476, 0.6895]$ & $[8.7824, 9.5990]$ & $(5.0188, 0.6686)$ & 9.1849 \\
\hline
1.92 & $[1.91, 1.93]$ & $[4.8303, 5.0963]$ & $[0.6410, 0.6823]$ & $[8.6416, 9.4327]$ & $(4.9633, 0.6616)$ & 9.0316 \\
\hline
1.94 & $[1.93, 1.95]$ & $[4.7793, 5.0388]$ & $[0.6346, 0.6752]$ & $[8.5053, 9.2721]$ & $(4.9090, 0.6549)$ & 8.8834 \\
\hline
\end{tabular}
\end{table*}
Fig.~\ref{fig:bifbox} presents the numerically computed loci of saddle-node bifurcations associated with the smaller values of $\alpha_1$ on the $(n_1,\alpha_1)$ parameter slice. Our certified enclosures use the fixed parameter values $(\alpha_2,\delta_1,\delta_2,n_2)=(12,1.2,0.8,1.6)$ and vary $(\alpha_1,n_1)$ over the reported ranges. For context, we also plot a second numerical fold locus for an alternative fixed parameter set (set B) to illustrate sensitivity of the fold location to the fixed parameters. The interval-certified saddle-node candidate enclosures from Table~\ref{tab:recommended_slice_sn_strips} are overlaid for reference. The numerical fold loci serve as a non-validated baseline (obtained via pointwise multistart Newton solves at discrete parameter values on the corresponding slice) against which the certified enclosures can be visually compared. Fig.~\ref{fig:smaller_larger_alpha1_folds} complements this view by showing, for the fixed parameter set used in certification, both the smaller- and larger-$\alpha_1$ numerical fold loci and the interval between them.
\subsubsection{Time-domain validation across the candidate strips.}
We ran simulations on both sides of the certified saddle-node candidate region (Fig.~\ref{fig:bifbox}, Table~\ref{tab:recommended_slice_sn_strips}). The observed change in long-term behavior is consistent with a transition across a saddle-node candidate boundary between monostable and bistable regimes.
\subsection{Certified Hopf candidate enclosures}
\begin{figure}[h!]
\centering
\includegraphics[width=0.75\columnwidth]{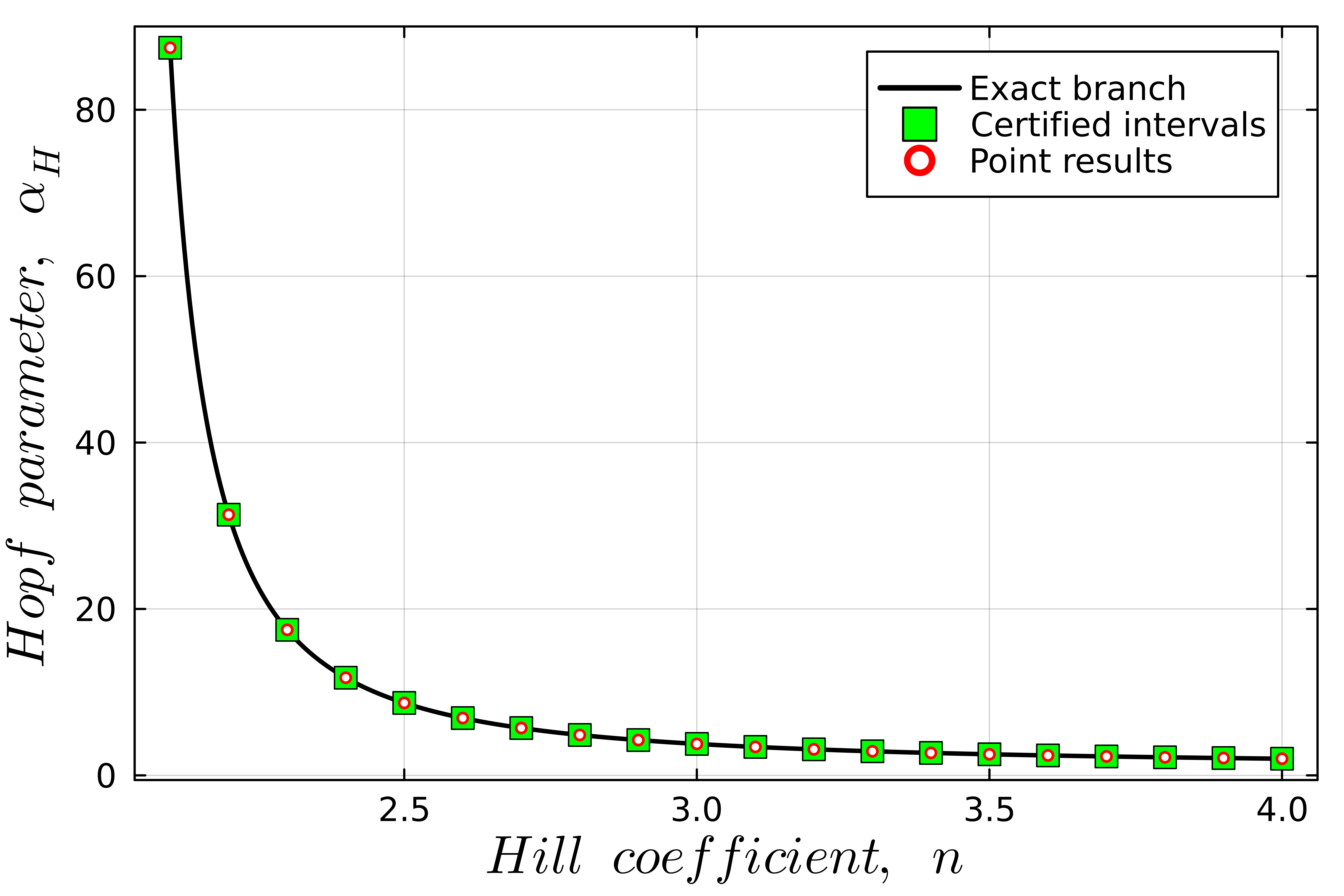}
\caption{Repressilator-type oscillator (symmetric branch): projection of validated Hopf \emph{candidate} enclosures onto the $(n,\alpha_H)$ slice. Intervals summarize candidate $\alpha_H$ values consistent with Table~\ref{tab:hopf_repressilator}. The \emph{exact branch} is the analytic symmetric equilibrium branch ($x_1=x_2=x_3=x^*$) from the equilibrium equations. Each enclosure also satisfies strict Routh--Hurwitz sign conditions $a_1>0$, $a_2>0$, and $a_3>0$ (interval-verified).}
\label{fig:hopf_boundary}
\end{figure}

We consider the symmetric repressilator model
\begin{equation}
\begin{aligned}
  \dot{x}_1&=\frac{\alpha}{1+x_3^n}-x_1;\,
  \dot{x}_2&=\frac{\alpha}{1+x_1^n}-x_2;\,
  \dot{x}_3&=\frac{\alpha}{1+x_2^n}-x_3,
\end{aligned}
\label{eq:repressilator}
\end{equation}
and restrict attention to the symmetric equilibrium branch $x_1=x_2=x_3=x^*$. On this branch, Hopf candidates can be characterized by a three-variable real system in $(x^*,\alpha,\omega)$; we certify solutions on the $(n,\alpha)$ slice by applying Algorithm~\ref{alg:ibf} with the three-state Routh--Hurwitz Hopf conditions in \eqref{eq:hopf_rh}.
For each Hill coefficient value $n$, the procedure returns validated enclosures of the equilibrium $x^*$, the Hopf candidate parameter $\alpha_{H}$ (here $\alpha_H\equiv\alpha$), and the oscillation frequency $\omega_{H}$. For each accepted enclosure, we additionally verify the strict inequalities $a_1>0$, $a_2>0$, and $a_3>0$ by checking that the corresponding interval enclosures are strictly positive. Certified results are reported in Table~\ref{tab:hopf_repressilator} (representative subset due to space constraints), and the corresponding candidate boundary samples in the selected slice are summarized in Fig.~\ref{fig:hopf_boundary}.

\begin{table}[h!]
\centering
\caption{Repressilator-type oscillator case study (symmetric branch): locally validated Hopf \emph{candidate} enclosures obtained using the three-state Routh--Hurwitz condition \eqref{eq:hopf_rh}. Each row reports validated interval bounds for the equilibrium value $x^*$ (with $x_1=x_2=x_3=x^*$), the associated candidate parameter $\alpha_H$, and the oscillation frequency $\omega_H$ at the sampled Hill coefficient $n$.}
\label{tab:hopf_repressilator}
\scriptsize
\resizebox{\columnwidth}{!}{%
\begin{tabular}{|c|c|c|c|}
\hline
\rowcolor{blue!10}
\textbf{$n$} & \textbf{$x^*$} & \textbf{$\alpha_H$} & \textbf{$\omega_H$} \\
\hline
2.5 & [1.741098, 1.741104] & [8.705457, 8.705554] & [1.732051, 1.732051] \\
\hline
2.6 & [1.588936, 1.588941] & [6.885371, 6.885433] & [1.732051, 1.732051] \\
\hline
2.7 & [1.475241, 1.475245] & [5.690203, 5.690247] & [1.732051, 1.732051] \\
\hline
2.8 & [1.387142, 1.387145] & [4.854987, 4.855019] & [1.732051, 1.732051] \\
\hline
2.9 & [1.316987, 1.31699] & [4.243616, 4.243641] & [1.732051, 1.732051] \\
\hline
3 & [1.25992, 1.259922] & [3.779753, 3.779773] & [1.732051, 1.732051] \\
\hline
3.1 & [1.212701, 1.212703] & [3.417606, 3.417622] & [1.732051, 1.732051] \\
\hline
3.2 & [1.173079, 1.173081] & [3.128208, 3.128221] & [1.732051, 1.732051] \\
\hline
3.3 & [1.139443, 1.139445] & [2.892429, 2.89244] & [1.732051, 1.732051] \\
\hline
3.4 & [1.110604, 1.110605] & [2.697178, 2.697187] & [1.732051, 1.732051] \\
\hline
3.5 & [1.085667, 1.085668] & [2.53322, 2.533228] & [1.732051, 1.732051] \\
\hline
\end{tabular}
}
\end{table}

\section{Discussion}\label{sec:discussion}
Rather than returning a single bifurcation point at a nominal parameter value, the proposed method returns collections of interval enclosures. Their projections summarize regions in parameter space that are certified to contain (or to exclude) solutions of the saddle-node or Hopf \emph{candidate} equations. Each accepted box provides a proof of existence and local uniqueness of a candidate solution within that box, while excluded boxes provide certified absence for the same candidate system. For the three-state oscillator, restricting to the symmetric branch and applying the Routh--Hurwitz reduction yields analogous certificates for Hopf candidates without introducing eigenvector variables.

Computational performance is largely governed by interval overestimation in the augmented equations, particularly in Jacobian- and determinant-based quantities, which can weaken Krawczyk pruning and lead to deeper bisection. Performance improves with structure-exploiting boxes (e.g., positivity, symmetry, conservation) and warm-starting across neighboring slice parameters.
The reported certifications are intentionally local: they validate candidates within the tested boxes but do not, by themselves, establish that all candidates on a slice have been exhaustively located. Slice-level absence/presence statements require domain-covering hybrid search with explicit bookkeeping of certified, excluded, and undecided regions, together with problem-specific contractors tailored to the underlying kinetics.

Future work includes (i) validated checks of genericity and transversality so that certified algebraic candidates can be upgraded to certified saddle-node/Hopf bifurcations and (ii) slice-level searches that provide both candidate enclosures and certified exclusions over larger regions. For Hopf points, an additional step is a validated computation of the first Lyapunov coefficient using interval-enclosed higher-order derivatives.

\section{Conclusions}\label{sec:conclusion}
This study introduces a rigorously validated interval-analysis workflow for certifying saddle-node and Hopf \emph{candidate} conditions in nonlinear ordinary differential equation models subject to bounded parametric uncertainty. By applying Krawczyk-type inclusion--exclusion tests to square, augmented candidate systems, the method yields box-local certificates that either (i) guarantee the existence and local uniqueness of a candidate solution or (ii) rigorously certify its absence. Numerical case studies demonstrate the effectiveness of the framework by producing certified saddle-node candidate enclosures for a bistable two-state circuit and certified Hopf candidate enclosures for a three-state oscillator via a Routh--Hurwitz-based specialization. Overall, the proposed workflow offers a mathematically rigorous and computationally tractable mechanism for certifying regime transitions under parametric uncertainty, thereby complementing sampling-based analyses and non-validated bifurcation-detection baselines (e.g., numerical continuation or pointwise multistart Newton solves at discrete parameter values) with formally justified certificates. Beyond synthetic gene networks, the workflow is applicable to interval-uncertain nonlinear ODE models in chemical reaction networks and biochemical signaling circuits, and to oscillatory models in power systems and power electronics. These certificates are most directly useful when parameters are only known within bounded intervals (e.g., after estimation), and one needs guaranteed regime-boundary localization/exclusion to support robust design, safety/feasibility screening, or experiment selection near qualitative transitions.

\bibliography{ifacconf} 
\end{document}